\documentclass[letterpaper,english,aps,letterpaper,english,reprint]{revtex4-1}
\usepackage{lmodern}
\usepackage{lmodern}
\usepackage[latin9]{inputenc}
\usepackage{textcomp}
\usepackage{amsmath}
\usepackage{amssymb}
\usepackage{graphicx}
\usepackage{esint}

\makeatletter

\usepackage{mathrsfs}

\usepackage{babel}

\usepackage{babel}

\usepackage{babel}

\makeatother

\usepackage{babel}
\begin{document}

\title{Spin-wave beam propagation in ferromagnetic thin film with graded
refractive index: mirage effect and prospective applications}

\author{Pawel Gruszecki and Maciej Krawczyk}

\affiliation{Faculty of Physics, Adam Mickiewicz University in Poznan, Umultowska
85, Pozna\'{n}, 61-614, Poland}

\date{\today}
\begin{abstract}
Using analysis of iso-frequency contours of the spin-wave dispersion
relation, supported by micromagnetic simulations, we study the propagation
of spin-wave (SW) beams in thin ferromagnetic films through the areas
of the inhomogeneous refractive index. We compare the transmission
and reflection of SWs in areas with gradual and step variation of
the SW refractive index. In particular, we show the mirage effect
for SWs with narrowing SW beam width, and an application of the gradual
modulation of the SWs refractive index as a diverging lens. Furthermore,
we study the propagation of SWs in ferromagnetic stripe with modulated
refractive index. We demonstrate that the system can be considered
as the graded-index waveguide, which preserves the width of the SW
beam for a long distance \textendash{} the property essential for
prospective applications of magnonics. 
\end{abstract}
\maketitle

\section{Introduction}

Spin-waves (SWs) are promising information carriers considered for
efficient and low energy consuming information processing devices\textendash magnonic
units, being able to supplement or even replace standard CMOS circuits
\cite{bernstein2010device,nikonov2013overview,krawczyk2014review,chumak2014magnon}.
However, before practical utilization of SWs, methods for efficient
excitation, transduction, and control of propagating SWs in nanoscale
planar structures need to be developed. Although SWs are characterized
by complex dispersion relation, many phenomena and practical solutions
 known from photonics can be exploited and transferred to magnonics
\cite{demidov2008mode,stigloher2016snell}. In photonics, control
of light propagation with the design of spatially varied refractive
index has found broad spectra of applications, ranged from fibers
to metamaterials \cite{saleh1991fundamentals,cai2009optical,solymar2009waves}.
Especially profitable are graded-index (GI) materials, i.e., materials
with a gradual change of the refractive index  \cite{hecht1998optics}.
Design of the refractive index topography enables manipulation of
the direction, velocity, and phase of the propagating waves. A naturally
occurring optical phenomenon related to the gradual decrease of the
refractive index is the \textit{mirage} \cite{hecht1998optics}. This
effect takes place when light bends near a warmed-up region (e.g.,
a ground or a road), where, due to a gradient of the air temperature,
the gradual decrease of the refractive index occurs. A well-known
example of the mirage is a \textit{Fata Morgana}. In fiber communication,
additional dielectric cladding to the core is used to improve transmission
properties. It protects the transmitted signal from leaking energy
by reducing the influence of any roughness and irregularities of the
outer surfaces of the fiber. The refractive index between cladding
and core region can be changed either step-like or continuously, providing
step-index and GI fibers, respectively. Usage of GI fibers reduces
modal dispersion and significantly improves the efficiency of signal
transmission, especially in multimode fibers \cite{hecht1998optics,senior2009optical}.

In magnonics, there are many ways to modulate SW RI. That can be done
by modification of materials properties, such as the saturation magnetization,
the exchange stiffness, or the magnetic anisotropy, but also by structural
design (geometrical pattern), a change of the magnetic field magnitude
or the magnetic configuration. All these properties and related refractive
index values can be varied in a continuous way. Very good example
of such non-uniformity is the demagnetizing field naturally existing
at the edges of ferromagnetic films \cite{gruszecki2014goos,gruszecki2015influence,perez2015magnetic}
or a noncollinear magnetization \cite{davies2015towards,xing2016fiber,yu2016magnetic}.
The gradual change of the refractive index can be also introduced
during device fabrication by nanostructuralization, ion implantation,
or voltage \cite{kakizakai2017influence}. Furthermore, it is possible
to modulate refractive index dynamically by a change of the external
magnetic field, e.g., using a magnetic field generated by DC current
\cite{houshang2016spin,demokritov2004tunneling,hansen2007resonant,ahmed2017guided},
the voltage across the film \cite{kakizakai2017influence}, or temperature
\cite{vogel2015optically,busse2015scenario}.

The influence of non-uniformity of the static external magnetic field
on propagating SWs has already been studied. However, the normal incidence
of magnetostatic SWs onto a region with a perturbed profile of the
static external magnetic field with collinearly \cite{demokritov2004tunneling,hansen2007resonant,kostylev2007resonant,neumann2009frequency}
or noncolinearly \cite{hauser2016yttrium} magnetized thin films has
been considered. Recently, we have reported an investigation of SW
beam reflection from the vicinity of the interface with gradual refractive
index due to the demagnetizing field \cite{gruszecki2014goos,gruszecki2015influence}.
Also, the SW propagation in noncollinear magnetization has been exploited
to demonstrate GI magnonics as a promising field of research for utilization
\cite{davies2015towards}.

A prospective application of magnetic media with the gradual change
of the refractive index in magnonics is the guiding of SWs. In the
recent theoretical papers guiding along the domain walls was considered
\cite{xing2016fiber,wagner2016magnetic}, also the confinement in
the region between domain walls with chirality appearing due to presence
of the Dzyaloshinskii-Moriya interaction was studied \cite{yu2016magnetic,garcia2015narrow}.
Nonetheless, an oblique incidence of SWs onto a region with a gradual
change of magnetic properties in ferromagnetic films and GI magnonic
waveguides have not yet been extensively explored \cite{zubkov1999trajectories,zubkov2007magnetostatic,vashkovsky1990passage},
and we contribute to this field in this paper.

In the paper, using iso-frequency dispersion contours analysis \cite{lock2008properties}
in order to develop ray optics approximation for SWs, supported by
micromagnetic simulations, we study the SW beam propagation in thin
ferromagnetic films and waveguides, which are made from thin yttrium
iron garnet (YIG) film. YIG is a dielectric magnetic material highly
suitable for magnonic applications due to its low damping \cite{serga2010yig}.
Recently, fabrication of very thin YIG films with thicknesses down
to tens of nanometers, preserving low damping \cite{sun2012growth,sun2013damping,hauser2016yttrium,krysztofik2017characterization},
which can be patterned in nanoscale \cite{liu2014ferromagnetic,pirro2014spin,krysztofik2017Damping}
has been demonstrated. For the sake of simplicity, our attention is
concentrated on the investigation of thin YIG films, out-of-plane
(OOP) magnetized by the external magnetic field. The change of the
refractive index is obtained by variation of the magnitude of the
static effective magnetic field. Nonetheless, the model can be extended
for an in-plane magnetized film, after taking into account proper
dispersion relation. The analytical predictions are validated by micromagnetic
simulations. In particular, we show, that with a decrease (increase)
of the internal magnetic field value $H$, the SW refractive index
increases (decreases). We define conditions for total internal reflection
and show how that phenomenon depends on $\mathrm{grad}H$. Interestingly,
for a slow increase of $H$ in space, we observe a mirage effect for
SWs. For a rapid change of $H$ value, we get the significant lateral
shift of the SW beam along the interface. Moreover, comparison of
the results obtained for gradual and step changes of the refractive
index in the ferromagnetic stripe suggests, that GI waveguides can
offer improved transmission of the SW beam.

The paper is organized as follows. In Sec.~\ref{sec:2Model-and-methods}
we present the analytical model and the micromagnetic simulations.
The obtained results for the extended thin YIG film and stripe waveguides
are discussed in Sec.~\ref{sec:Results}. Conclusions are provided
in Sec.~\ref{sec:Summary}.

\section{Model and methods\label{sec:2Model-and-methods}}

\subsection{Spin-wave dynamics\label{subsec:2_1Model}}

We consider a thin ferromagnetic film with the thickness ($L_{z}$)
much smaller than the lateral dimensions of the film ($L_{z}\ll L_{x},L_{y}$).
The film is saturated with the static external magnetic field $\mathbf{H}$.
Magnetization dynamics is described by the Landau-Lifshitz-Gilbert
(LLG) equation of motion for the magnetization vector $\mathbf{M}$
\cite{gilbert2004phenomenological}: 
\begin{equation}
\frac{\mathrm{d}\mathbf{M}}{\mathrm{d}t}=-\frac{\left|\gamma\right|\mu_{0}}{1+\alpha^{2}}\mathbf{M}\times\mathbf{H}_{\mathrm{eff}}-\frac{\alpha\left|\gamma\right|\mu_{0}}{M_{\mathrm{S}}\left(1+\alpha^{2}\right)}\mathbf{M}\times\left(\mathbf{M}\times\mathbf{H}_{\mathrm{eff}}\right),\label{eq:LLE}
\end{equation}
where $\alpha$ is the damping parameter, $\gamma$ is the gyromagnetic
ratio, $\mathbf{H}_{\mathrm{eff}}$ is the effective magnetic field,
$M_{\mathrm{S}}$ is magnetization saturation. The first term in the
LLG equation describes the precessional motion of the magnetization
around the effective magnetic field and the second term enriches that
precession by damping. The effective magnetic field, in general, can
consist of many terms. In this paper we consider the contributions
of the external magnetic field, the exchange field $\mathbf{H}_{\mathrm{ex}}$
and the dipolar field $\mathbf{H}_{\mathrm{d}}$: $\mathbf{H}_{\mathrm{eff}}=\mathbf{H}+\mathbf{H}_{\mathrm{ex}}+\mathbf{H}_{\mathrm{d}}$.

In the case of OOP uniformly magnetized thin film the SW dispersion
relation is given by \cite{kalinikos1986theory,stancil2009spin}:
\begin{equation}
\omega^{2}=\left(\omega_{\mathrm{H}}+l_{\mathrm{ex}}^{2}\omega_{\mathrm{M}}k^{2}\right)\left(\omega_{\mathrm{H}}+l_{\mathrm{ex}}^{2}\omega_{\mathrm{M}}k^{2}+\omega_{\mathrm{M}}F\left(kL_{z}\right)\right)\label{eq:DR_OOP}
\end{equation}
where: $\omega=2\pi f$ is the angular frequency of SWs, $f$ is the
frequency, $\mu_{0}$ is the permeability of vacuum, $\omega_{\mathrm{H}}=\left|\gamma\right|\mu_{0}(H-M_{\mathrm{S}})$,
$\omega_{\mathrm{M}}=\gamma\mu_{0}M_{\mathrm{S}}$, and the exchange
length $l_{\mathrm{ex}}=\sqrt{2A/(\mu_{0}M_{\mathrm{S}}^{2})}$, $A$
is the exchange constant, $k$ is the wave number and 
\begin{equation}
F\left(x\right)=1-\frac{\left(1-\text{e}^{-x}\right)}{x}.\label{eq:fl-1-1}
\end{equation}
If $kL_{z}\gg1$, the dispersion relation can be simplified to the
following equation: 
\begin{equation}
\omega^{2}=\left(\omega_{\mathrm{H}}+l_{\mathrm{ex}}^{2}\omega_{\mathrm{M}}k^{2}\right)\left(\omega_{\mathrm{H}}+l_{\mathrm{ex}}^{2}\omega_{\mathrm{M}}k^{2}+\omega_{\mathrm{M}}\right).
\end{equation}

It is clear from Eq.~(\ref{eq:DR_OOP}), that the SW dispersion relation
is isotropic for any frequency in the OOP magnetized film. This makes
the analysis simpler and we will study only OOP magnetized thin films
and stripes in this paper. In the case of in-plane magnetized films
at low frequencies, where dipolar contribution dominates, the dispersion
relation is anisotropic. Nevertheless, with increasing frequency,
the iso-frequency contours smoothly transform through elliptical to
almost circular at high frequencies, where SW dynamic is determined
by the exchange interactions \cite{zubkov2007magnetostatic,lock2008properties}.
Thus, the implementation of the analytical model developed below for
OOP configuration to the in-plane magnetized thin films can be done
by taking the analytical dispersion relation for SW in the in-plane
magnetized films from Ref.~\cite{kalinikos1986theory} instead of
Eq.~(\ref{eq:DR_OOP}).

\begin{figure}[!th]
\includegraphics[width=8.6cm]{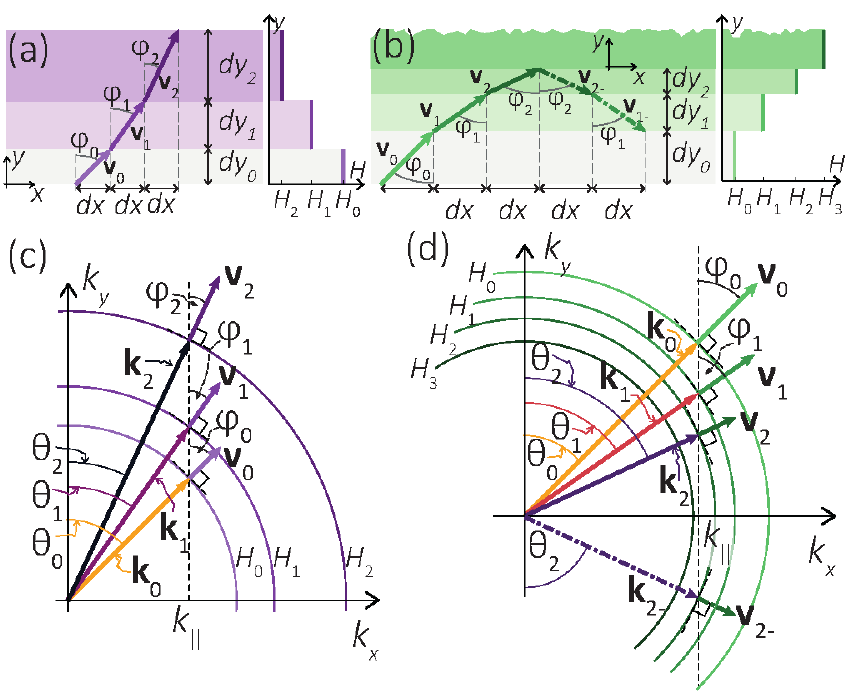} \protect\protect\caption{Schematic representation of the gradual bending of obliquely incident
SW onto the region with increased {[}(a), (c){]}, and decreased {[}(b),
(d){]} value of the external magnetic field $H$. The alternate iso-frequency
contours in (c) and (d) correspond to the alternate regions of decreasing
and increasing values of the internal magnetic field $H_{i}$ in (a)
and (b), respectively. The wave vectors $\textbf{k}_{i}$, vectors
of the group velocities $\textbf{v}_{i}$, and their angles with respect
to the $y$-axis, $\theta_{i}$ and $\varphi_{i}$, respectively,
are indicated for each magnetic field segment $i$. The vertical dashed
lines in (c) and (d) mark the conservation of the tangential component
of the wave vector $k_{\|}$ at successive refractions and reflection.
The insets on the right side of (a) and (b) correspond to $H$ values
in accordingly marked regions.\label{fig:F2_AnModel} }
\end{figure}

\subsection{Ray optic approximation\label{subsec:Methods_IFC}}

Let us analyze SWs propagation in a medium with slow, but a stepwise
change of the internal magnetic field value along the $y$-axis from
$H_{0}$, as shown schematically in the insets in Fig.~\ref{fig:F2_AnModel}
(a) and (b). We assume, that the dispersion relation obtained for
a uniform thin film magnetized by the homogeneous in space external
magnetic field (Eq.~\ref{eq:DR_OOP}) can be used in each segment
$i$ of the constant magnetic field $H_{i}$. SW ray propagating through
an area with a gradual change of the internal magnetic field will
be gradually bent due to the change of the magnetic field resulting
in the variation of the SWs refractive index  {[}see Fig.~\ref{fig:F2_AnModel}
(c) and (d){]}.

The bending of SW ray can be estimated from the conservation of the
tangential to the interface component of the wave vector $k_{\|}$,
$k_{i,\|}=k_{i+1,\|}$ \cite{stigloher2016snell}. As the interface,
we refer to the $xz$ planes being perpendicular to the gradient of
the magnetic field, which separates two successive segments with different
magnetic fields, $H_{i}$, and $H_{i+1}$. Schematically it is shown
in Fig.~\ref{fig:F2_AnModel}(c) and (d). To model SW beam propagation
we will consider SW ray which can be treated as a curve on which are
laying centroids of the SW beam. Rays follow the changes of the SWs
group velocity direction ($\mathbf{v}_{i}$) as it is shown in Fig.~\ref{fig:F2_AnModel}(a)
and (b) where $H$ increases and decreases along the $y$-axis, respectively.
If we mark the angle of SW beam propagation by $\varphi$ (the angle
of ($\mathbf{v}_{i}$) with respect to the $y$-axis) we can write
\begin{equation}
dy=\cot\varphi(y)dx,
\end{equation}
and the function defining SW ray path $y\left(x\right)$ can be expressed
in the recursive form:

\begin{equation}
y_{N}=y_{N-1}+\cot\varphi\left(y_{N-1}\right)dx\label{eq:SWRay}
\end{equation}
where $y_{N}=y\left(x_{N}\right)$, $x_{N}=x_{0}+Ndx$ and SW starts
propagation from the point $\left(x_{0},y_{0}\right)$ with the angle
of incidence $\varphi_{0}\equiv\varphi(y_{0})$. The direction of
propagation, $\varphi$, is related to the direction of the energy
transfer, i.e., the direction of the group velocity vector ($\mathbf{v}_{\mathrm{g}}=\nabla_{\mathbf{k}}\omega(\mathbf{k})$,
where $\omega(\mathbf{k})=\omega(k_{x},k_{y})$ is a dispersion relation).
The group velocity is normal to the iso-frequency contours. In the
case of isotropic dispersion (considered in the paper), the iso-frequency
contours are circular and the direction of the group and phase velocities
are equal $\varphi=\theta$ where $\theta=\cot^{-1}k_{y}/k_{x}$.
The angle of incidence can be expressed as 
\begin{equation}
\cot\varphi(k_{x},k_{y})=\frac{v_{\mathrm{g},y}}{v_{\mathrm{g},x}}=\frac{\frac{\partial}{\partial k_{y}}\omega\left(k_{x},k_{y}\right)}{\frac{\partial}{\partial k_{x}}\omega\left(k_{x},k_{y}\right)}\label{eq:cot_varPHI}
\end{equation}
where $v_{\mathrm{g},n}$ is $n$-th component of the group velocity.
The initial conditions for the incident SW are known, hence a value
of $k_{x}$ and angular frequency, $\omega$ at the starting point
of SW are known. The normal component of the wavevector $k_{y}\left(\omega,H\right)$,
can be calculated from Eq.~(\ref{eq:DR_OOP}) numerically or analytically.
Exemplary iso-frequency contours lines for different values of the
internal magnetic field $H_{i}$ with marked directions of group velocities
corresponding to the fixed value $k_{\|}$ are shown in Fig.~\ref{fig:F2_AnModel}
(c) and (d).

Apart from refraction at the interface, there is possible also reflection.
If the variation of the magnetic field between successive segments
is small, most of the SW energy is transmitted (in the analytical
model transmission is considered only), unless total reflection condition
is fulfilled. In total reflection, there are no available solutions
corresponding to the given $k_{\|}$ of the incident wave at some
value of the internal magnetic field, e.g., in Fig.~\ref{fig:F2_AnModel}
(d) $|\textbf{k}_{3}|<k_{2,\|}$. In such a case SW is reflected from
the interface, see Fig.~\ref{fig:F2_AnModel}(b) and (d) where such
a situation is presented at the interface between $H_{2}$ and $H_{3}$.
According to the law of reflection $\theta_{\mathrm{inc}}=\theta_{\mathrm{ref}}$
and $k_{y,\mathrm{ref}}=-k_{y,\mathrm{inc}}$, where superscripts
``inc'' and ``ref'' refer to the incident and reflected waves,
respectively. Taking into account both, the bending and reflection
of SWs, we can predict the SWs ray path for the conserved tangential
component of the wavevector to the interface (aligned along the $x$-axis,
$k_{x}=\mathrm{const.}$) using the following procedure:

\begin{equation}
dy=\begin{cases}
\cot\varphi(y)dx & \mathrm{if\,}k\left(\omega,H\left(y\right)\right)\geq k_{x}\\
-\cot\varphi(y-dy)dx & \mathrm{if\,}k\left(\omega,H\left(y\right)\right)<k_{x}.
\end{cases}\label{eq:reflection_cond}
\end{equation}

A similar model of the SWs propagation in a medium with a gradual
change of the $\mathbf{H}$ caused by the demagnetizing field induced
in the vicinity of the film's edge was presented in Ref.~\cite{gruszecki2015influence}.
However, that model was valid only for isotropic SWs dispersion. This
limitation is removed here due to the analysis of the group velocity
direction instead of the wave vector.

\subsection{Micromagnetic simulations}

Micromagnetic simulations have been proven to be an efficient tool
for the calculation of SW dynamics in ferromagnetic materials. Presented
results were obtained using the MuMax3 \cite{vansteenkiste2014design}
which solves time-dependent LLG equation (\ref{eq:LLE}) with included
Landau damping term with the finite difference method. In simulations,
we consider an oblique SW beam propagation in YIG thin film saturated
by an OOP magnetic field. We assume typical magnetic parameters of
YIG at 0K, it is \textbf{$A=0.4\times10^{-11}$} J/m, $M_{\mathrm{S}}=0.194\times10^{6}$
A/m, $\gamma=176$ rad GHz/T and the value of damping $\alpha=0.0005$.
The system of size $L_{x}\times L_{y}\times L_{z}$ was discretized
with cuboid elements of dimensions $l_{x}\times l_{y}\times L_{z}$.
Lateral dimensions of the single cell $l_{x}\times l_{y}$ and film
thickness $L_{z}=10$ nm are less than the exchange length of YIG,
$13$~nm. The simulations have been performed for two geometries:
i) 6 $\mu$m $\times$ 4 $\mu$m $\times$ 10 nm discretized with
the cell of lateral dimensions $2\times2$ nm$^{2}$ for high-frequency
exchange SWs and ii) 32 $\mu$m$\times$4 $\mu$m$\times$10 nm discretized
with the cell of size $8\times8$ nm$^{2}$ for SWs of lower frequency
(15 GHz).

Every simulation comprises two parts. First, we get the equilibrium
static magnetic configuration, which in our study is always OOP magnetization.
Then, the results of the first stage are used in the dynamic part
of simulations, which are aimed at obtaining the steady-state. SWs
are continuously generated in the form of a Gaussian beam which propagates
through the film. At the edges of the film $x=0$ and $x=L_{x}$,
absorbing boundary conditions are applied \footnote{The absorbing boundary has been implemented in form of a parabolic
increase of the damping value in the vicinity of the edges of the
simulated area (at the distance of ca. 5-10 wavelengths the damping
value increases up to 0.5). For exchange SWs additional absorbing
boundary conditions at $y=0$ and $y=L_{y}$ has been also applied.
See further details in Ref.~\cite{gruszecki2014goos,venkat2018absorbing}.}. SW beams are excited by means of the spatially non-uniform dynamic
external magnetic field at a given frequency and the spatial profile
designed to excite the SW beam of appropriate width. The profile of
the dynamic magnetic field used to excite the SW beam is similar to
the profile generated by a coplanar waveguide with modulated width
\cite{gruszecki2016microwave} multiplied by the Gaussian function
changing its value along the axis of the coplanar waveguide. The exact
description of the SWs' beam excitation can be found in Ref.~\cite{gruszecki2017goos}.
After sufficiently long time of continuous excitation, when the beam
is clearly visible and doesn't change qualitatively in time, a steady-state
is achieved. The data necessary for further analysis are stored. From
the stored micromagnetic simulations  results rays corresponding to
the excited SW beam are extracted. Firstly, the time average SW intensity
colormaps are obtained for simulation according to the equation: $I(x,y)=\frac{f}{4}\int_{0}^{4/f}\left[m_{x}\left(x,y,t\right)\right]^{2}dt$,
where $m_{x}$ is normalized component of the magnetization vector.
Then, the Gaussian fitting is applied to get ray line coordinates
(details can be found in Ref.~\cite{gruszecki2014goos}). Those ray
lines are directly compared with the results of the analytical model.

\begin{figure}[h!]
\includegraphics[width=8.6cm]{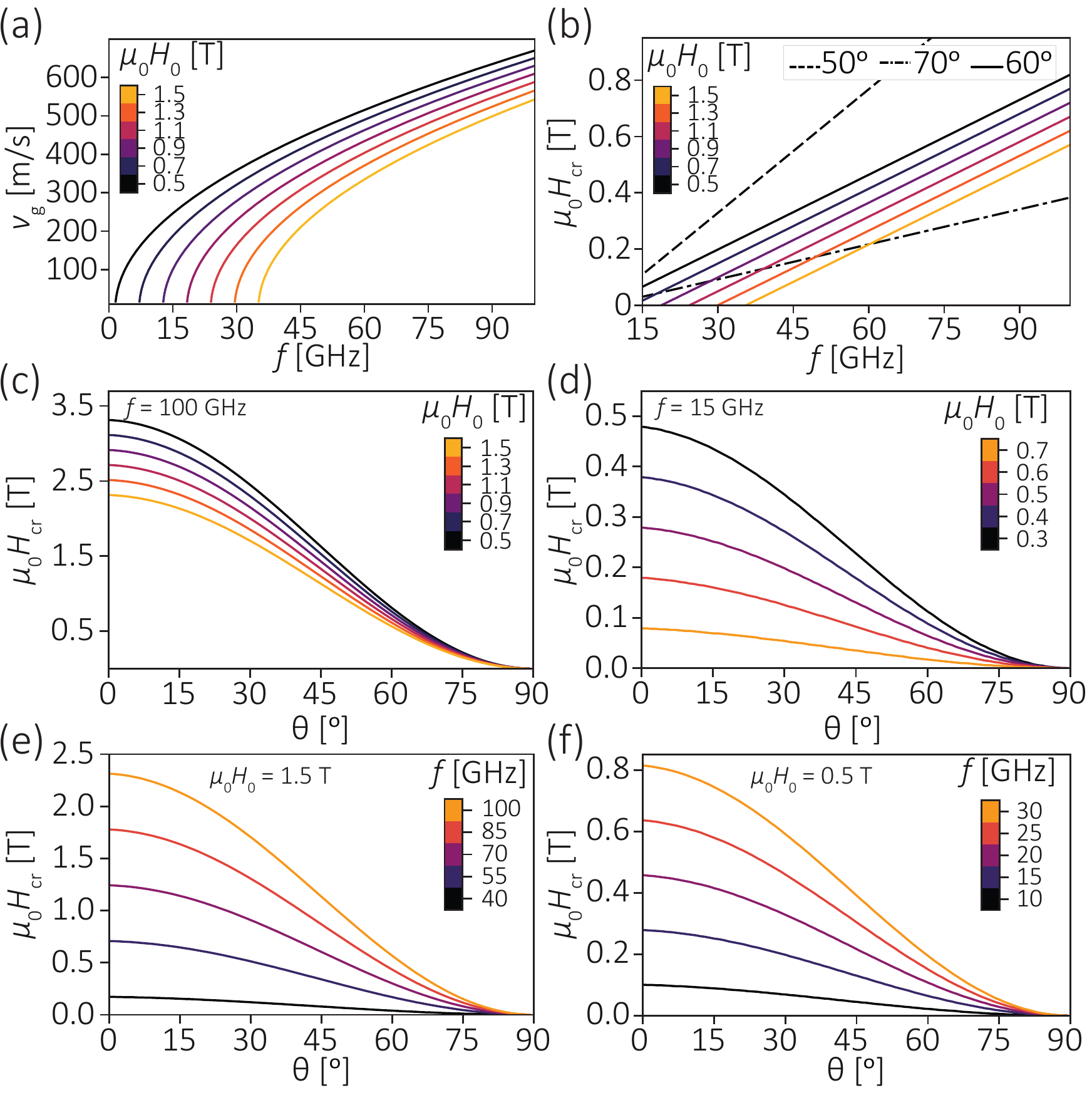} \protect\caption{(a) The group velocity of SWs in the OOP magnetized YIG film of thickness
10 nm in dependence on frequency, plotted for several values of the
homogeneous, static external magnetic field, $H_{0}$. (b) Critical
external magnetic field $H_{\mathrm{cr}}$ in dependence on SWs frequency
for the angle of incidence 60\textdegree{} and several values of the
external magnetic field (solid lines), and also for $\mu_{0}H_{0}=0.5$
T and the angle of incidence 50\textdegree{} (black dash-dotted line),
and 70\textdegree{} (black dashed line). $H_{\mathrm{cr}}$ in dependence
on the angle of incidence and several values of $H_{0}$ for SWs of
frequency, in (c) for 100 GHz and in (d) for 15 GHz. $H_{\mathrm{cr}}$
in dependence on the angle of incidence and several values of SWs
frequency for (e) $\mu_{0}H_{0}=1.5$ T and (f) $\mu_{0}H_{0}=0.5$
T.\label{fig:figR_AN}}
\end{figure}

\section{Results\label{sec:Results}}

\begin{figure*}
\includegraphics[width=18cm]{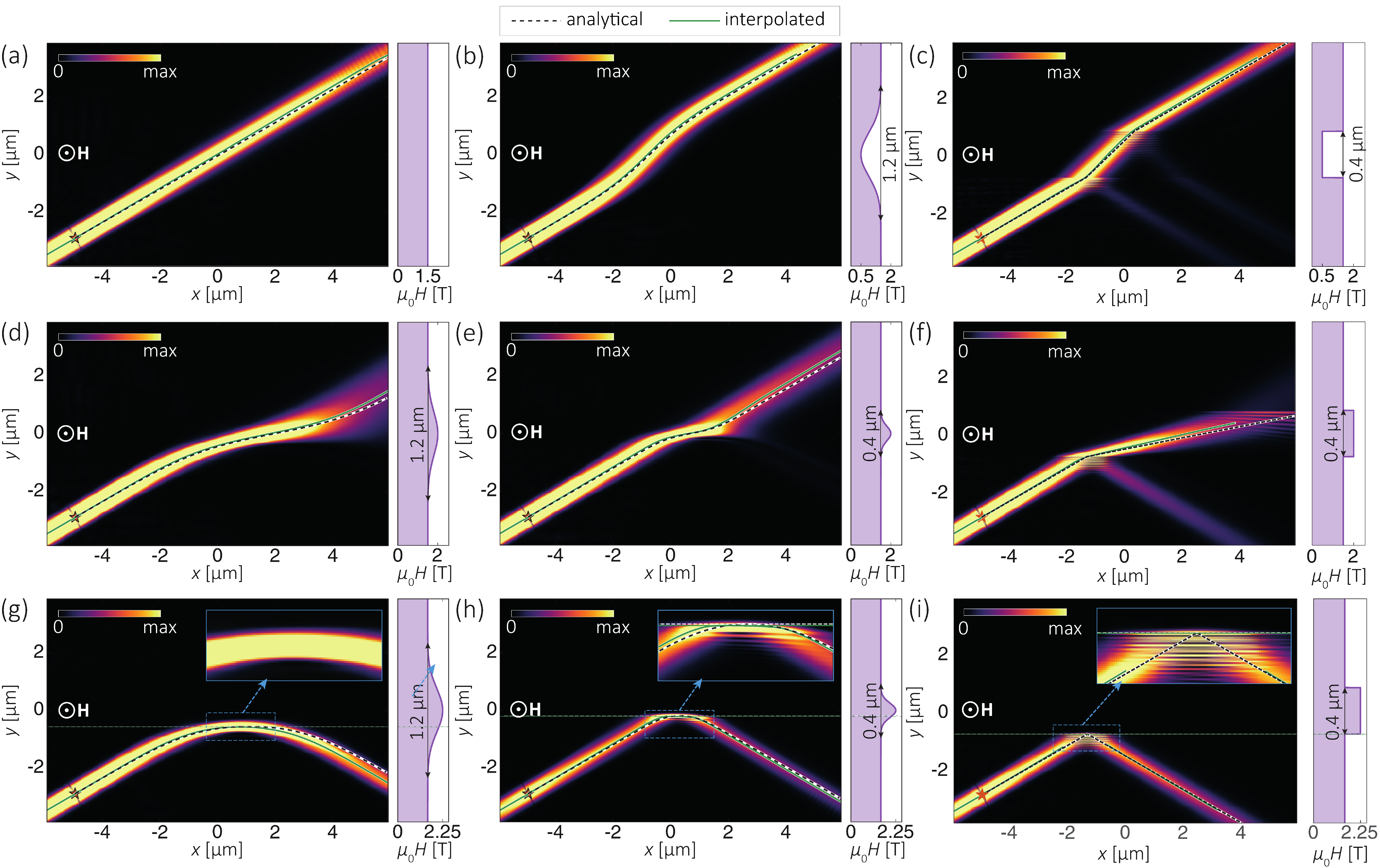} \protect\caption{The micromagnetic simulations  results presenting SWs intensity maps
obtained for SW beams of frequency $f=100$ GHz incident under the
angle $\theta=60$\textdegree{} in the $10$ nm thick YIG film. Solid
green lines and dashed black lines correspond to the results interpolated
from simulations and calculated using the analytical model, respectively.
The static magnetic field (directed OOP) depends only on the $y$
coordinate. Its profile is presented in the insets on the right side
of each figure. In (g)-(i), the horizontal overlapping lines correspond
to the plane of reflection and their positions obtained from the analytical
model and extracted from micromagnetic simulations  perfectly agree.
The insets in right top corners of (h) and (i) present zoomed in regions
of the simulated area marked via dashed green rectangle. (a) SW beam
propagation in the medium with uniform $H$; incline SW beam propagation
through the region with the (b) gradual and (c) step decrease of the
magnetic field. (d) and (e) gradually increased $H$ value up to 2.0
T at $y=0$ with a different gradient of $H$ changes; (f) step-index
change of $H$ up to 2.0 T in the central region of the film. (g)
and (h) gradually increased $H$ value up to 2.25 T for $y=0$ with
different profiles of $H$; (i) step-index change of $H$ up to 2.25
T in the central region of the film. In (d)-(e) SW beam is spread
due to small field increase, this $H$ modulation effectively works
as an diverging lens; in (e) the small part of SW beam energy is also
reflected; in (f) a part of the energy is reflected at both first
and second interface. In (g)-(i) the total internal reflection of
SW beam is visible. (g) The total internal reflections at the region
with the gradually changing refractive index   is referred to as a
mirage. In (h) is visible a large shift between the incident and reflected
beam spots, at some range, SW beam propagates parallel to the interface.
\label{fig:figSIM}}
\end{figure*}

\subsection{Analytical model}

%\subsubsection*{Spin-wave propagation in OOP magnetized film}

In OOP magnetized thin film the SW ray angle with respect to the $y$-axis
is given by 
\begin{equation}
\cot\varphi(y)=\cot\theta(y)=\frac{k_{y}}{k_{x}}=\frac{\sqrt{k^{2}\left(\omega,H\left(y\right)\right)-k_{x}^{2}}}{k_{x}},
\end{equation}
where the value of $k\left(\omega,H\left(y\right)\right)$ can be
calculated numerically from the dispersion relation Eq.~(\ref{eq:DR_OOP}).
The results of the analytical analysis of the SW rays are shown in
Fig.~\ref{fig:F2_AnModel}(a) and (b) for the increased and decreased
internal magnetic field, respectively.

%\subsubsection*{SWs velocity}

The knowledge about the group velocity of SWs during their propagation
through the film is important from an application point of view. Exemplary
results showing how $v_{\mathrm{g}}$ depends on the SWs frequency
for different values of the external magnetic field are presented
in Fig.~\ref{fig:figR_AN}(a). It is shown, that $v_{\mathrm{g}}$
for OOP configuration monotonously increases (apart from the region
of dominating magnetostatic interactions at low frequencies, invisible
in the figure)\cite{stancil2009spin} with an increase of the frequency
and decrease of the external magnetic field value. Therefore, SWs
falling at a region with decreased $H$, and thus the increased refractive
index, accelerates. That situation is opposite to optics, where an
increase of the refractive index is related to a decrease of the group
velocity of electromagnetic waves.

%\subsubsection*{Total internal reflection}

The effect of total internal reflection is important for wave applications,
especially in designing of the waveguides. We will analyze the total
internal reflection of SWs in the OOP magnetized film with spatially
modulated magnitude of the static external magnetic field $H(y)=H_{0}+H'(y)$,
where $H'(y)$ is its modulation, which in the case under investigation,
depends only on the $y$ coordinate. Therefore, we will analyze the
critical field $H_{\mathrm{cr}}$ at which the total internal reflection
takes place in dependence on frequency, $H_{0}$, and the angle of
incidence. The results for $\mu_{0}H_{0}=$ 0.5-1.5 T with an interval
0.2 T and $\varphi=60\text{\textdegree}$, and additionally, for angles
of incidence 50\textdegree{} and 70\textdegree{} at $\mu_{0}H_{0}=0.5$
T are shown in Fig.~\ref{fig:figR_AN}(b). It is visible that with
the increase of $\varphi$ the smaller value of the field $H_{\mathrm{cr}}$
is required to obtain total internal reflection. Furthermore, the
greater $\varphi$ the smaller slope of $H_{\mathrm{cr}}\left(f\right)$,
and interestingly, these dependencies are linear.

The dependencies of $H_{\mathrm{cr}}$ on the angle of incidence for
different values of $H_{0}$ and $f$ are in details presented in
Fig.~\ref{fig:figR_AN}(c)-(f). In Fig.~\ref{fig:figR_AN}(c) and
(d) are dependencies of $H_{\mathrm{cr}}$ on the angle of incidence
for SWs at frequencies 100 GHz and 15 GHz, respectively, and for different
values of $H_{0}$. It is visible, that while the angle of incidence
increases, the value of $H_{\mathrm{cr}}$ decreases. Moreover, the
higher value of $H_{0}$ the smaller $H_{\mathrm{cr}}$ is needed
to obtain total internal reflection. It is, because, while $H$ increases
the FMR (ferromagnetic resonance) frequency increases as well. Therefore,
the greater length of wavevector (frequency is much higher than the
FMR frequency) the higher value of $H_{\mathrm{cr}}$ is required
to obtain total internal reflection. In Fig.~\ref{fig:figR_AN}(e)
and (f) are presented dependencies of $H_{\mathrm{cr}}$ on the angle
of incidence for $\mu_{0}H_{0}=$ 1.5 and 0.5 T, respectively, for
different values of $f$. It is shown, that the higher frequency,
the larger modulation of $H'(y)$ is needed to obtain total internal
reflection, what is consistent with previous analysis.

\begin{figure*}
\includegraphics[width=18cm]{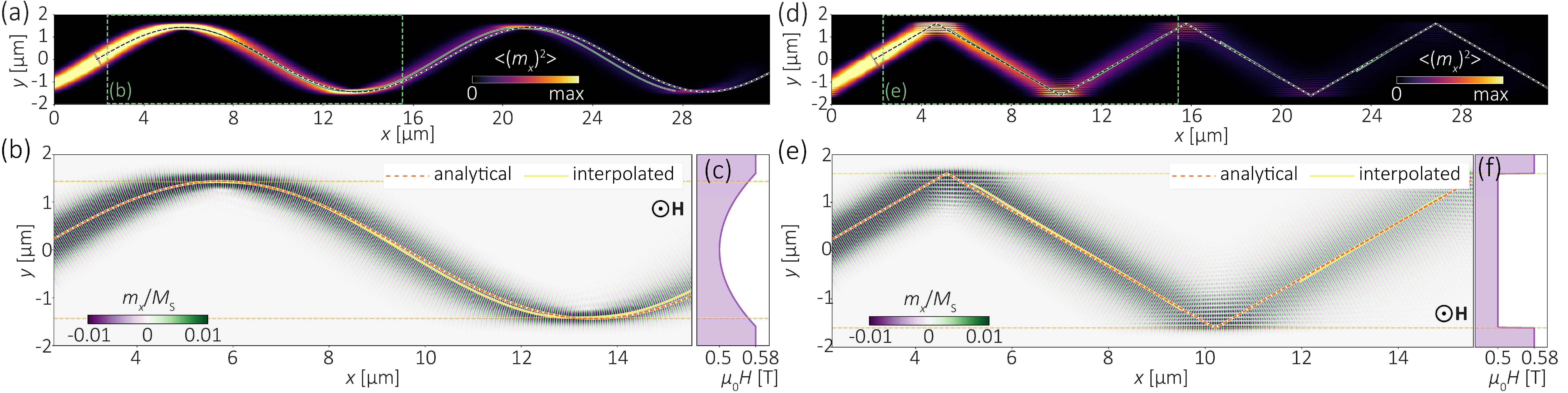} \protect\caption{Spin-wave beam propagating under the angle of 60\textdegree{} at frequency
$f=15$ GHz in $10$ nm thick and 4 $\mu$m wide YIG stripe in presence
of gradual (a)-(c) and step (d)-(f) changes of the external magnetic
field across the stripe's widths. The respective profiles of the magnetic
field across the stripe are shown in (c) and (f). In (a) and (d) the
SWs intensity obtained from micromagnetic simulations  is shown; in
(b) and (e) the amplitude of SWs from the fragment of the structures
presented in (a) and (d) (marked by the green rectangle with dashed
sides), respectively, are plotted. The dashed red lines show the result
of the ray model which match well with the ray extracted from micromagnetic
simulations, marked with the solid-yellow lines. \label{fig:figWideWG}}
\end{figure*}

\subsection{Micromagnetic simulations}

The predictions of the analytical model we validated by means of micromagnetic
simulations. Those simulations were performed for i) high-frequency
$f=100$ GHz exchange dominated SWs and ii) lower frequency SWs, $f=15$
GHz.

%\subsubsection*{Exchange spin-waves}

In Fig.~\ref{fig:figSIM} we show the SW intensity maps (the square
of the dynamic component of the magnetization averaged over time:
$\left\langle m_{x}^{2}\left(x,y,t\right)\right\rangle _{1/f}$).
The results are presented for the exchange dominated SW beams of the
beam width c.a. 700 nm and wavelength 33.6 nm (at 100 GHz) incident
under the angle 60\textdegree{} at the region with varied magnetic
field $H\left(y\right)=H_{0}+H'\left(y\right)$, where $\mu_{0}H_{0}=1.5$
T and $H'\left(y\right)$ is a Gaussian function. Additionally, we
show SW rays obtained from the analytical model (black-dashed lines)
from Eq.~(\ref{eq:reflection_cond}) and extracted from the MS results
(green-solid line) for different profiles of $H'\left(y\right)$ in
(a)-(i). Overall, the very good agreement between micromagnetic simulations
 and the analytical model is found.

Accordingly with the model predictions in Fig.~\ref{fig:F2_AnModel}(a)
the decrease of $H$ results in the gradual decrease of the angle
of refraction, see Fig.~\ref{fig:figSIM}(b) and (c). It causes the
shift of the transmitted beam along the $x$-axis with respect to
the case of unbent SW beam {[}Fig.~\ref{fig:figSIM}(a){]}. In the
case of gradually decreasing $H\left(y\right)$ the SW beam reflections
from the nonuniform region\textbf{ }are not visible\textbf{ }{[}Fig.~\ref{fig:figSIM}(b){]},
whereas in the case of step-index change of $H\left(y\right)$ the
reflected SW beams are apparent {[}Fig.~\ref{fig:figSIM}(c){]}.
In the case of step-index change of $H\left(y\right)$, the interference
pattern of the incident and reflected waves is visible as horizontal
stripes with higher and lower intensities near the interface.

The increase of $H(y)$ causes the increase of the angle of refraction
{[}Fig.~\ref{fig:figSIM}(d)-(i){]}, which is also according to the
model estimation shown in Fig.~\ref{fig:F2_AnModel}(a). When the
increase of the magnetic field is slow and the maximal value of $H\left(y\right)$
is smaller than $H_{0}+H_{\mathrm{cr}}$, the refracted SW beam is
spread, see Fig.~\ref{fig:figSIM}(d). Hence, that region with increased
$H$ can be treated as a diverging lens for SWs in analogy to optics
\cite{hecht1998optics}. For more abrupt changes of $H\left(y\right)$,
clearly visible reflected SW beams appears {[}Fig.~\ref{fig:figSIM}(e){]}.
In the limit of the step-index change of $H\left(y\right)$ the incident
SW beam is split into transmitted and pronounced reflected SW beams,
see Fig.~\ref{fig:figSIM}(f).

For higher maximal values of the external magnetic field, exceeding
the condition for total internal reflection $\max\left(H\left(y\right)\right)>H_{0}+H_{\mathrm{cr}}$,
the incident beam is totally reflected {[}Fig.~\ref{fig:figSIM}(g)-(i){]}.
It is noteworthy that the magnetic field at which the total internal
reflection takes place is almost identical in the case of micromagnetic
simulations and analytical model, see two overlapping horizontal dashed
lines in Fig.~\ref{fig:figSIM}(g)-(i). These horizontal lines correspond
to the value of the external magnetic field $H=H_{0}+H_{\mathrm{cr}}$
and can be referred to as the interface at which total internal reflection
takes place. Interestingly, for the slow increase of $H(y)$, the
equivalent of the mirage effect for SWs is apparent {[}see Fig.~\ref{fig:figSIM}(g){]}.
It means, that the wavefronts of incident SW beam are gradually bent
without reflection, and the interference pattern near the interface
is not visible.

For the rapid change of the $H(y)$ magnitude, the interference pattern
near the area of the varied magnetic field is present, pointing at
the reflection of waves. In this scenario, the SW ray near the interface
becomes almost parallel to its line for some distance, see Fig.~\ref{fig:figSIM}(h).
It means, that between the incident and reflected SW beam spots appears
lateral shift along the interface of value almost 1 $\mu$m. One may
call this phenomenon as Goos-Hanchen effect for SWs \cite{gruszecki2014goos,goos1947neuer}
of surprisingly high value. However, we fully recreated this effect
using ray optics approximation only (see the dashed black line), whereas
the literature Goos-Hanchen effect is a wave phenomenon, defined as
a lateral shift between the incident and reflected beam spots due
to phase change occurring at the interface \cite{artmann1948berechnung}.
It means, the Goos-Hanchen effect cannot be explained by the use of
ray optics approximation, therefore, the observed lateral shift shouldn't
be referred to as the Goos-Hanchen shift, although, the observed phenomenon
looks equivalently in the far field. For the step change of $H(y)$
{[}Fig.~\ref{fig:figSIM}(i){]}, neither lateral shift nor bending
are visible. The interference pattern is clearly apparent and there
is no transmission to the upper part of the sample.

%\subsubsection*{Exchange-dipolar spin-waves}

In the last part of our study, we verify analytical predictions for
lower frequency SWs and the ferromagnetic stripe of the finite width.
We analyze the SW beam at the frequency $f=15$ GHz (wavelength 115
nm) with the beam width 680 nm propagating in the 10 nm thick and
4 $\mu$m wide YIG stripe OOP magnetized by the external magnetic
field. The value of $H(y)$ in the middle part is set as $\mu_{0}H_{0}=0.5$
T and its magnitude gradually increases when moving to the stripe
edges up to $\mu_{0}H\left(w\right)=0.58$ T near the sides of the
stripe, at a distance $w=1.6$ $\mu$m from the center of the stripe
{[}see Fig.~\ref{fig:figWideWG}(c){]}. The quadratic change of the
field $H\left(y\right)=H_{0}+H_{\mathrm{m}}^{\prime}y^{2}/w^{2}$,
where $\mu_{0}H_{\mathrm{m}}^{\prime}=0.08$ T, is assumed.

In Fig.~\ref{fig:figWideWG}(a) the SW beam propagates under the
angle 60\textdegree{} with respect to the $y$-axis counted in the
middle part of stripe and is multiple times reflected in the considered
part of the waveguide. Under the assumption of the realistic value
of the damping in YIG, the SW beam propagates for a distance up to
30 $\mu$m with reasonable intensity. The bending of the wavefronts
in the region with the gradually increased magnetic field is demonstrated
in Fig.~\ref{fig:figWideWG}(b), which is similar to the observation
made in Fig.~\ref{fig:figSIM}(g) for high-frequency SWs. The ray
of the propagating SW beam obtained from the analytical model, Eq.~(\ref{eq:reflection_cond}),
is shown with the red dashed line. The satisfactory agreement between
analytical and simulation results is found at the beginning part of
the waveguide, but this negligible discrepancy, increases with a distance,
pointing out that wave effects, which are not taken into account in
the ray model, exist in the propagation through GI media, and accumulate
with a distance.

\begin{figure}
\includegraphics[width=8.6cm]{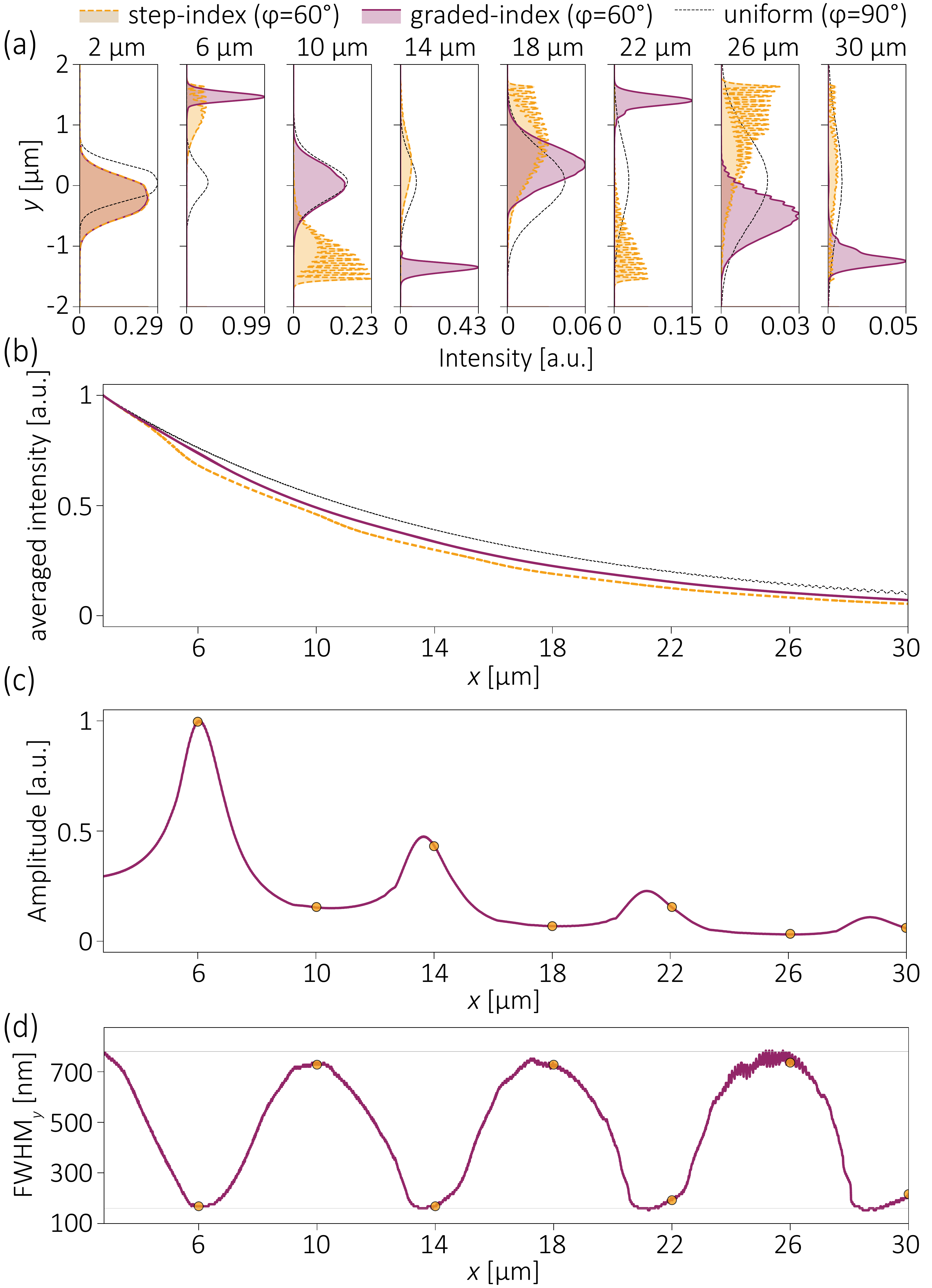} \protect\caption{(a) Spin-wave beam profile across the width of stripe presented in
Fig.~\ref{fig:figWideWG} for several values of $x$. Solid violet
and dashed yellow lines have been obtained for the SW beam profiles
excited with $\varphi=60^{\circ}$ and propagating in the GI and the
step index stripes, respectively. The thin black dashed line corresponds
to the SW beam of the same width but propagating along the $x$-axis
($\varphi=90^{\circ})$ in the uniform magnetic field. The SWs intensity
is normalized to the maximal intensity in the whole waveguide for
all simulations. (b) The averaged intensity of SWs across the stripe's
width as the function of $x$ for two different profiles of the external
magnetic field: step-index (dash-dotted magenta line), and GI (dashed
yellow line). These results are compared with SW beam propagating
along the $x$-axis in the uniformly magnetized stripe. (c) Maximal
amplitude of the SW beam in GI stripe in dependence on $x$. (d) SW
beam's width (full width at half maximum) evolution in the GI stripe
in dependence on the $x$ coordinate. Golden dots in (c) and (d) correspond
to the profiles of SW beam in GI stripe presented in (a). \label{fig:figPropLength-1}}
\end{figure}

For a comparison, micromagnetic simulations  have been performed for
the step-index change of $H\left(y\right)$ at $y=\pm w$, see the
field profile in Fig.~\ref{fig:figWideWG}(f). The SW beam in such
a system {[}see Fig.~\ref{fig:figWideWG}(d) and (e){]} can propagate
in the form of the narrow beam for a shorter distance, than in the
GI waveguide. The additional simulation has been performed also for
the SW beam of the same profile in the beam waist propagating along
the $x$-axis ($\varphi=90^{\circ}$) in the waveguide saturated by
the uniform external magnetic field of value $\mu_{0}H_{0}=0.5$ T.
The results of simulations are presented in Fig.~\ref{fig:figPropLength-1}(a),
where profiles of intensities of SW beams propagating in above described
three waveguides (with GI, step-index and constant profile of the
magnetic field) at eight different distances along the $x$-axis are
shown. It is clear that the SW beam propagating in the GI waveguide
preserves its width for a much longer distance than other two waveguides.
For instance, at $x=26$ $\mu$m the SW beam width is comparable with
that at $x=2$ $\mu$m, whereas for step-index waveguide SW beam is
spread across the whole width of the waveguide. Interestingly, the
SW beam in GI waveguide is much narrower, than the input beam, in
the areas where the total internal reflection occurs (see profile
for $x=6$ $\mu$m).

Furthermore, we analyze the change of the averaged intensity of SW
beams along the waveguide's width with increasing $x$ for different
$H\left(y\right)$ profiles, see Fig.~\ref{fig:figPropLength-1}
(b). For all profiles of $H\left(y\right)$ the dependencies are similar
because it is due to damping constant $\alpha$ in Eq.~\ref{eq:LLE}.
One can observe, that decay of the SW beam amplitde excited with $\varphi=90^{\circ}$
is slightly slower than for beams excited with $\varphi=60^{\circ}$
in the GI and step-index waveguides. However, the difference results
from different optical paths of the beams in the analyzed examples,
i.e., the shortest path is obtained for the SW beam excited with $\varphi=90^{\circ}$
in a waveguide magnetized by the uniform field $H(y)=H_{0}$, and
the longest for the SW beam excited with $\varphi=60^{\circ}$ propagating
in the step-index fiber. SW beam propagating in the GI waveguide,
due to its bending, possess the moderate effective length of the beam
ray.

Finally, we analyze in details SW beam behavior in GI waveguide. In
Fig.~\ref{fig:figPropLength-1}(c) and (d) the SW beam amplitude
and the full width at half maximum (FWHM$_{y}$) taken along the $y$-axis,
respectively, are plotted in dependence on $x$. The local maxima
in Fig.~\ref{fig:figPropLength-1}(c) correspond to the SWs beam
narrowing while it propagates near the waveguide edges. In the region
near the critical value of $H$ SWs intensity is around 3 times larger
than in the central part of the waveguide. Therefore, the more pronounced
maxima, the more collimated SW beam is. This is confirmed by the FWHM$_{y}\left(x\right)$,
which oscillates between its maximal width (located near the center
of the waveguide, FWHM$_{y}=780$ nm) and the minimal width (located
near the critical value of the $H$, FWHM$_{y}=160$ nm). Any systematic
beam's spreading along the $x$ is visible and the ratio between maximal
and minimal widths is kept being equal 4. Overall, it means that the
stripe with GI modulation of $H\left(y\right)$ preserves the best
transmission properties from all analyzed cases.

The significant difference in SW beam propagation between the stripe
with the gradual and step change of the static magnetic field, proofs
that the stripes with graded refractive index are promising candidates
for transmitting signals carried by SWs. The system under investigation
can be referred to as a wide, multimode GI waveguide for SWs. We believe
the proposed system can be a good playground to study SW beam guiding
and can be directly used for an experimental demonstration of the
investigated effects, e.g., using recently experimentally validated
in Ref.~\cite{korner2017excitation} method of SW beams excitation.

%If the variation of the magnetic field between sucessive segments is small, most of the SW energy will be transmitted, unless total reflection condition will be fulfilled.

\section{Summary\label{sec:Summary}}

We have studied SW beam propagation in the ferromagnetic film and
waveguide with a gradual change of the magnetic field. These structures
can be referred to as magnonic GI media, for which non-uniformity
of the refractive index is introduced by the variation of the external
magnetic field magnitude. For the sake of simplicity, we performed
all investigations for the out-of-plane magnetized thin YIG films.
Nevertheless, the study can be extended into other materials and other
magnetic configurations. We have proposed the ray optic approximation
to describe SW beam propagation through an area of the ferromagnetic
film with the gradual variation of the magnetic field value. We have
successfully verified the analytical model by micromagnetic simulations
for high and relatively low-frequency SWs in an exchange regime, for
which the magnetostatic effects are negligible. We have demonstrated
bending of the SW beam propagating obliquely through regions with
GI, the total internal reflection, and the mirage effect with narrowing
of the SW beam. We have also demonstrated that a region with the inhomogeneous
magnetic field value can be used to obtain a diverging SW lens. Thin
ferromagnetic stripe with an inhomogeneous magnetic field across its
width has been used to study GI materials for SW guiding. The obtained
results have been compared with results for the stripe with a step-index
change of the external magnetic field. We show, that the stripe with
the gradual modulation of the SW refractive index possesses better
transmission properties than the step-index stripe. Interestingly,
SW beam propagating in the GI stripe is periodically narrowed in the
region of the reflection at areas with the internal field fulfilling
the condition for total internal reflection, and then again spreads.
Ultimately, any systematic beam spreading has not been observed for
stripe with GI, as opposed to the step-index stripe, or SW beam propagating
along the stripes axis. It points out, that this approach can be used
to transmit SW beam without beam widening for very long distances,
limited only by damping. We believe, that the use of SWs waveguides
with additional GI cladding shall significantly reduce the influence
of defects at the edges of the ferromagnetic stripes, reduce spreading
of the SW beam width, and enhance the transmission. Further investigations
are necessary for the optimization of the multi- and single-mode SW
waveguides, and for other magnetization configurations.
\begin{acknowledgments}
This project has received funding from the European Union's Horizon
2020 research and innovation programme under the Marie Sk\l odowska-Curie
grant agreement No 644348 and National Science Centre of Poland project
UMO-2012/07/E/ST3/00538. 
\end{acknowledgments}

\bibliographystyle{apsrev4-1}
\addcontentsline{toc}{section}{\refname}\bibliography{examplebib,literature}

\end{document}